\documentstyle[12pt,epsf,epsfig,amsmath]{article}
\def\eqright #1\cr{\noalign{\hfill$\displaystyle{{}#1}$}}
\def\eqleft #1\cr{\noalign{\noindent$\displaystyle{{}#1}$\hfill}}

\def\oldreffmt#1{\rlap{[#1]} \hbox to 2\parindent{}}

\def\figfmt#1{\rlap{Figure {#1}} \hbox to 1in{}}

\def\sectioneq{\def\theequation{\thesection.\arabic{equation}}{\let
\holdsection=\section\def\section{\setcounter{equation}{0}\holdsection}}}%

\newcounter{holdequation}

\def\begineq #1\endeq{$$ \refstepcounter{equation}\eqalign{#1}\eqno
	(\theequation) $$}
\def\contlimit{\,{\hbox{$\longrightarrow$}\kern-1.8em\lower1ex
\hbox{${\scriptstyle (a\rightarrow0)}$}}\,}
\def\centeron#1#2{{\setbox0=\hbox{#1}\setbox1=\hbox{#2}\ifdim
\wd1>\wd0\kern.5\wd1\kern-.5\wd0\fi
\copy0\kern-.5\wd0\kern-.5\wd1\copy1\ifdim\wd0>\wd1
\kern.5\wd0\kern-.5\wd1\fi}}
\def\centerover#1#2{\centeron{#1}{\setbox0=\hbox{#1}\setbox
1=\hbox{#2}\raise\ht0\hbox{\raise\dp1\hbox{\copy1}}}}
\def\centerunder#1#2{\centeron{#1}{\setbox0=\hbox{#1}\setbox
1=\hbox{#2}\lower\dp0\hbox{\lower\ht1\hbox{\copy1}}}}
\def\lsim{\;\centeron{\raise.35ex\hbox{$<$}}{\lower.65ex\hbox
{$\sim$}}\;}
\def\gsim{\;\centeron{\raise.35ex\hbox{$>$}}{\lower.65ex\hbox
{$\sim$}}\;}

\def\super#1{\ifmmode \hbox{\textsuper{#1}}\else\textsuper{#1}\fi}
\def\textsuper#1{\newcount\holdspacefactor\holdspacefactor=\spacefactor
$^{#1}$\spacefactor=\holdspacefactor}

\def\getcite#1,{\advance\citenumber by1
\def\getcitearg{#1}\def\lastarg{@}
\ifnum\citenumber=1
\ref{#1}\let\next=\getcite\else\ifx\getcitearg\lastarg\let\next=\relax
\else ,\ref{#1}\let\next=\getcite\fi\fi\next}

\def\pom{{\rm P\kern -0.53em\llap I\,}}
\def\spom{{\rm P\kern -0.36em\llap \small I\,}}
\def\sspom{{\rm P\kern -0.33em\llap \footnotesize I\,}}

\relax
\def\contlimit{\,{\hbox{$\longrightarrow$}\kern-1.8em\lower1ex
\hbox{${\scriptstyle (a\rightarrow0)}$}}\,}
\def\upon #1/#2 {{\textstyle{#1\over #2}}}
\relax
\renewcommand{\thefootnote}{\fnsymbol{footnote}}

\sectioneq

\def\til#1{\centeron{\hbox{$#1$}}{\lower 2ex\hbox{$\char'176$}}}
\def\tild#1{\centeron{\hbox{$\,#1$}}{\lower 2.5ex\hbox{$\char'176$}}}
\def\sumtil{\centeron{\hbox{$\displaystyle\sum$}}{\lower
-1.5ex\hbox{$\widetilde{\phantom{xx}}$}}}

\begin{document} 

\begin{titlepage} 

$~$

\vspace{1in} 

\begin{center} 
  
{\large\bf Unification of the Standard Model Via High-Energy Unitarity}

\medskip

Alan. R. White\footnote{arw@hep.anl.gov }

\vskip 0.6cm

\centerline{Argonne National Laboratory}
\centerline{9700 South Cass, Il 60439, USA.}
\vspace{0.5cm}

\end{center}

\begin{abstract}
 
High-energy unitarity is satisfied by adding a sextet quark doublet to
QCD. The sextet sector produces electroweak symmetry breaking and 
is predicted to give large cross-section effects at the LHC. It 
embeds, uniquely, in a massless $SU(5)$ theory whose bound-state S-Matrix,
potentially, reproduces the Standard Model. 
Infra-red chirality transitions of the massless Dirac sea play an essential
dynamical role. 

\end{abstract} 

\vspace{2in}

\centerline{Contribution  to the 
Proceedings of HSQCD 2005 (recently requested).}

\renewcommand{\thefootnote}{\arabic{footnote}} \end{titlepage}

\section{Introduction} 

In this talk I will argue that the constraints of high-energy, multi-regge, unitarity 
are so strong that the Standard Model S-Matrix may be uniquely selected, 
together with an underlying,
unifying, massless SU(5) gauge theory (QUD\footnote{QUD $\leftrightarrow$ 
{\small \it Quantum Unodynamics, or
Quantum Unification Dynamics, or Quantum Unitary Dynamics}}). The key points are - 
\begin{enumerate}
\openup-1\jot {\it \item{The unitary Critical Pomeron uniquely 
selects QCD$_S~$ (a color sextet quark sector is added to the 
known triplet quarks) as the strong interaction.}
\item{Multi-regge scattering of electroweak bosons induces the 
QCD$_S$ bound-state 
S-Matrix via anomalies. Sextet chiral symmetry breaking produces vector boson masses.} 
\item{QCD$_S$ and the electroweak sector embed, uniquely, in SU(5) gauge theory 
with massless, left-handed, fermions in the representation $5+15+40+45^*~
\leftrightarrow$ QUD.  } 
\item{Remarkably, the QUD bound-state S-Matrix may
have only the interactions of the Standard Model,
with the known physical states as the low mass spectrum. }}
\end{enumerate}

The picture we develop appears to be consistent with all existing experimental
data. A crucial, and very unconventional, element 
is that the new physics producing electroweak symmetry breaking 
is due to a high mass sector of the (QCD) strong interaction that
is predicted\cite{arw051}, to produce  
dramatic, large cross-section, effects at the LHC. The new sextet sector also
provides a natural explanation\cite{arw051} for the major mysteries of dark matter and 
the cosmic ray spectrum knee, as well as
various other (currently mysterious) phenomena in both accelerator and
cosmic ray physics. 
 
We have arrived at QUD by singularly exploring research directions and problems 
considered ``too difficult'' and 
``far from the mainstream'', by the dictates of current fashion. 
As a result, our formalism appears obscure (for many physicists)
and radical with respect to the current theoretical paradigm, making 
our arguments hard to present and their credibility easily questioned. Of course, it would 
be incredible if the Standard Model, with all of it's complexity, has the underlying   
simplicity that we suggest. Nevertheless, all the needed ingredients are present
and, despite the enormous amount still to be established,  
everything points to the correctness of our suggestion.
If the predicted effects of the sextet sector
are seen at the LHC, interest in QUD will surely rise rapidly.

Both QCD$_S$ and QUD have special ultra-violet and infra-red properties 
that allow the construction of the high-energy, bound-state, S-Matrix 
using gauge theory reggeon diagrams and abstract multi-regge theory. 
Reggeons are gauge-invariant but carry the gauge group as a global
symmetry that is confined by reggeization infra-red divergences.  
Physical amplitudes are selected by a further divergence
due to chiral anomaly reggeon interactions\cite{arw051}.
With our regularization procedure, these interactions 
contain infra-red chirality transitions that, in effect, are the zero momentum
contribution of a propagator to a condensate. However, the chirality transitions are 
present only in the S-Matrix and then only in special reggeon 
vertices obtained via the regge-limit reduction of large-order loop diagrams to 
effective triangle diagrams. The chirality transitions are crucial in the formation of 
physical states and in producing ``wee gluon'' interactions, but they do not break the 
short-distance gauge invariance. The resultant
symmetry breaking and mass generation is a purely S-Matrix phenomenon  
- surely a radical element of our picture.
{\it There is no ``Higgs field'', rather
the masslessness of the field theory is essential,}

In the QCD$_S$ S-Matrix, there is confinement and chiral symmetry breaking but, 
in agreement with experiment, there is a much more limited spectrum\cite{arw051} than is 
conventionally expected in QCD. (There are no 
glueballs, no BFKL pomeron, and no odderon.)
When we first discovered\cite{kw} QUD, we were amazed and puzzled by the closeness 
of the triplet quark and lepton sectors 
to the Standard Model. (We asked only for the sextet sector.) With
the emergence, in the QUD multi-regge S-Matrix, of the SU(3)$\otimes$SU(2)$\otimes$U(1) 
interaction structure, 
the deeper significance of this feature becomes clear. The dynamics  
is analagous to that of QCD$_S$, but the gauge 
group appears broken at low transverse momentum 
because the anomaly divergences pick out (as the strong interaction) 
vector-like gauge boson exchanges invariant under an SU(3) subgroup.
SU(3) octet quarks with lepton-like quantum numbers 
also play a vital role. They do not couple to 
the strong interaction (pomeron), but their short-distance presence 
produces the generation structure of the physical states 
and also reconciles leptons with the underlying SU(5) symmetry.

In this talk, we will outline only how the basic structure of 
the Standard Model S-Matrix emerges from QUD. We will not discuss how 
the physical scales, and associated phenomena, appear.  
A paper, explaining the arguments in detail, and also elaborating the physics,
is in preparation\cite{arw07}.
We begin, very briefly, with the abstract formalism and 
unitarity constraints that provide the technical framework for our discussion.

\section{Reggeon Unitarity and the Critical Pomeron}

Multi-regge behavior is controlled by partial-wave 
amplitudes in which the (J-plane) singularity structure 
is determined by reggeon unitarity\cite{arw00} 
equations (obtained from multiparticle unitarity in the corresponding $t$-channel).
When first derived\cite{gpt} these equations 
were a spectacular generalization from low-order field theory
calculations. They were generally
accepted only after the development of multiparticle asymptotic
dispersion relations provided a fundamental basis for the complex angular-momentum
theory involved\cite{arw00}. 
Reggeon unitarity is satisfied by all existing gauge theory calculations
(including NLO BFKL) and plays a crucial role in our generalization of (still relatively
low-order) known results to the multi-regge region of both QCD$_S$ and QUD.

Reggeon Field Theory\cite{gr} (RFT) provides an ``effective field theory''
solution of reggeon unitarity. For a pomeron with intercept one, 
multi-pomeron singularities accumulate at $J=1$ (when $t=0$) but, 
beautifully, an RFT renormalization group (fixed-point) formalism can be 
applied to obtain an interacting pomeron theory with
the ``universality'' property of a critical 
phenomenon\cite{cri}. Scaling laws can be derived 
for many cross-sections and $s$-channel unitarity shown to be satisfied\cite{mm}.
The ``Critical Pomeron'' provides a complete, unique, 
solution of high-energy unitarity - with asymptotically rising cross-sections. 
Since no competitive solution exists (field-theoretic or otherwise)
the question is clearly whether there is a field theory 
which gives the Critical Pomeron.

\section{The Critical Pomeron and QCD$_S$}

The supercritical phase, in which the pomeron intercept is initially above one,
provides a direct link between the Critical Pomeron and QCD$_S$. 
A pomeron condensate pushes the physical intercept back below one
while also giving\cite{arw91} new classes of RFT diagrams 
in which a reggeized vector particle appears that
couples pairwise to the pomeron.
When SU(3) color is broken to SU(2) (giving ``color superconducting QCD'' 
- CSQCD) a single massive, reggeized, vector particle 
is deconfined - exactly as in the supercritical pomeron phase. A necessary condition
for critical behavior is that the scalar field involved be
asymptotically free, so that
SU(3) symmetry restores smoothly
at large and small
momentum. This requires\cite{gw} saturation of the QCD asymptotic freedom constraint\cite{arw051}. 
Also, we will see that a regge pole pomeron appears from
divergent reggeon diagrams only when there are scaling interactions due to 
the infra-red fixed-point produced by
the same saturation. The unique (physically realistic) possibility\cite{wm} to achieve saturation
is to add two color sextet quarks to the known
six triplets - giving QCD$_S$. The ``sextet pions''
can then become the longitudinal components of 
massive electroweak vector bosons. 

Because cross-sections fall in both the subcritical and supercritical phases,
while perturbative gauge theory cross-sections rise, 
if the pomeron is not critical it is unlikely that asymptotically free perturbation
theory can be matched with unitary forward amplitudes. Assuming
the explicit connection between 
CSQCD$_S$ and the supercritical pomeron is as described in the next Section, 
then QCD$_S$ uniquely gives the Critical Pomeron and unitarity of 
the strong interaction is linked directly to electroweak symmetry breaking. 

\section{The States and Amplitudes of QCD$_S$}
 
We consider multi-regge kinematic regions\cite{arw98} 
where, a priori, we expect\cite{arw051} to see the high-energy scattering of bound-state 
regge poles. We, initially, construct amplitudes with the 
reggeon diagrams of CSQCD$_S$
(as in the particular example shown in Figure 1). 
Reggeon unitarity determines that the reggeon diagrams 
involved are similar to elastic scattering
diagrams, except for the crucial difference that vertices coupling distinct
reggeon channels can contain triangle anomalies. In fact, an infra-red divergence 
selects, as bound-state physical amplitudes, those 
in which all reggeon channels are coupled via anomalies.

To describe the role of anomalies, 
we consider the multiple vector boson amplitude shown in Figure 1, in which there are color
zero massless gluon components in each reggeon channel. 
The divergence of reggeization
exponentiates to zero (in momentum space) all CSQCD$_S$ amplitudes with non-zero $SU(2)$
color in any channel. Figure 1 is the simplest amplitude in which
anomalies appear in all vertices. They are generated\cite{arw03,arw02} as illustrated in Figure 2, 
which also shows how an ``anomaly pole'' appears, via a chirality
transition,
when the gluon reggeons carry zero transverse momentum. (We discuss the deeper implications
of using vector boson external states in the next Section.) 
\begin{figure}[ht]
\centerline{\epsfxsize=4.5in \epsfbox{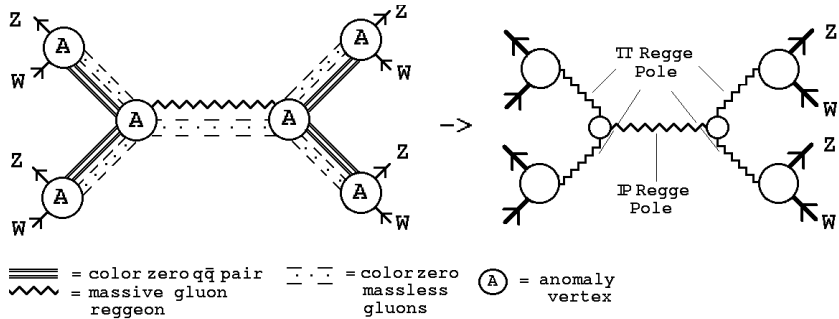}}   
\caption{Reggeon diagrams giving 
pion scattering via pomeron exchange.}
\end{figure}
\begin{figure}[ht]
\centerline{\epsfxsize=3.5in\epsfbox{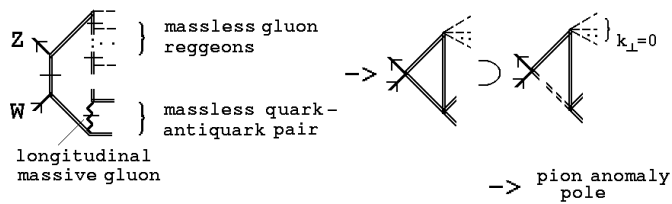}}   
\caption{Generation of a reggeon vertex triangle anomaly. The hatched
lines are on-shell and the broken quark line
indicates a zero momentum chirality transition.} 
\end{figure}

The initial effect of the anomalies 
is a large transverse momentum (non-unitary) power
enhancement\cite{arw03} of the high energy behavior. A 
cut-off removes the problem, but gauge invariation violation 
then gives an infra-red divergence as all
gluon transverse momenta scale 
to zero. If a massless gluon state has normal color 
parity ($= $ signature), interactions with additional reggeons
exponentiate this divergence. {\it It is not exponentiated only if all massless gluon 
states have ``anomalous'' ($\neq $ signature) color parity. In a vector theory, such states
couple only to anomalies.} Crucially, the fixed-point scaling of higher-order gluon interactions
preserves the ``anomalous wee gluon'' divergence to all orders.

The residue of the wee gluon 
divergence gives a physical CSQCD$_S$ amplitude in which the cut-off can be removed and all the
massless gluons in Figure 1 contribute only as an
{\it anomalous wee gluon condensate}. 
$SU(2)$ anomalous gluons have odd signature and so the
pomeron is an {\it even signature regge pole} which, because of the condensate, is
exchange degenerate with a reggeized massive gluon, just as in supercritical RFT.  
The anomaly poles produce\cite{arw051} {\it Goldstone boson} particles
that, because of the equivalence of conjugate
$SU(2)$ representations, include quark/quark and anti-quark/anti-quark {\it nucleons},
in addition to quark/antiquark {\it pions}. The condensate is a crucial 
component in both the particles and the pomeron. 
It provides the (vacuum equivalent) ``universal wee parton''
component of infinite momentum physical states. Within the anomaly vertex, 
the condensate is absorbed by the 
chirality transition of a zero momentum anti-quark (or quark), implying that 
a pion can be viewed as a pure 
quark/antiquark state, but either the quark or antiquark has to be in an unphysical, 
``negative energy'', state. 

{\it By removing the cut-off after the extraction of anomaly
infra-red divergences, we replace ultra-violet 
chirality violation (producing bad
high-energy behavior) by infra-red chirality violation  
producing particle poles. This is how a confining, chiral symmetry breaking, 
bound-state spectrum is generated out of perturbative reggeon diagrams.}

Assuming high-energy CSQCD$_S$ maps completely on to the supercritical pomeron 
(as all evidence suggests), then the transition to $QCD_S$ does indeed
give the Critical Pomeron. Because the triple pomeron vertex involves\cite{arw051}
a chirality transition, and the Critical Pomeron is an all-orders phenomenon,
there will be arbitrarily large numbers of transitions in any scattering process.
The wee gluon condensate, carrying fixed SU(3) color, will disappear and
instead there will be dynamical (effectively random) multi-reggeon 
gauge field fluctuations within the color group
(that are specifically allowed by
the Gribov ambiguity in the light-cone quantization of QCD.) The 
transition from a fixed ``magnetization'' for the Dirac sea shifting
gauge field, to a random, fluctuating, field is the 
``critical phenomenon'' underlying the Critical Pomeron. It provides a complex, but 
beautiful, wee parton (vacuum-like) phenomenon which makes a dramatic selection of 
the field theory degrees of freedom contributing to 
the S-Matrix. The gauge symmetry is not broken but color charge parity and
chiral symmetry are both broken spontaneously.

QCD$_S$ baryons are bound states of CSQCD$_S$ nucleons
and $SU(2)$ singlet quarks. (Most likely, 
the additional quark also contributes, initially, via a zero 
momentum chirality transition.) Because there are no chiral 
symmetries mixing the two sectors, there will be no ``hybrids'' consisting 
of sextet quarks (antiquarks) and triplet antiquarks (quarks). This has the very important
implication that the only new baryons will be 
the {\it sextet proton} and the {\it sextet neutron}. The neutron is expected to be stable
and so provides a naturally dominant (and very attractive) source of dark matter\cite{arw051}. 

For the physics of QCD$_S$ to be as we have described it,
the quarks have to be massless (implying many massless Goldstone bosons). 
This is a non-trivial problem, and it 
may very well be that the only possibility to introduce effective quark masses is via
the bound-state masses resulting from the embedding of $QCD_S$ in QUD discussed 
below.

\section{ QCD$_S$ and the Electroweak Sector }

The use of left-handed vector bosons as external states ensures that
our pions are Goldstone bosons of the weak interaction. Indeed, it could be that
the pion states and amplitudes we have described are not present in
QCD$_S$ in isolation and that the presence of the electroweak interaction is essential.
In fact, the nature of 
the physical states in QCD$_S$ is, in turn, essential for the generation of a mass
for an exchanged vector boson.

Sextet quarks (antiquarks) have the same $SU(3)$ triality as triplet antiquarks (quarks)
and so we anticipate that their electroweak couplings will be 
the same. At infinite momentum, the ``anomalous wee gluons'' (wee partons!) should
reproduce finite momentum ``vacuum properties''.
A wee gluon anomaly interaction generates\cite{arw051} 
a vector boson mass, via mixing with a pion anomaly pole, as illustrated in Figure 3.
\begin{figure}[ht]
\centerline{\epsfxsize=2.2in
\epsfbox{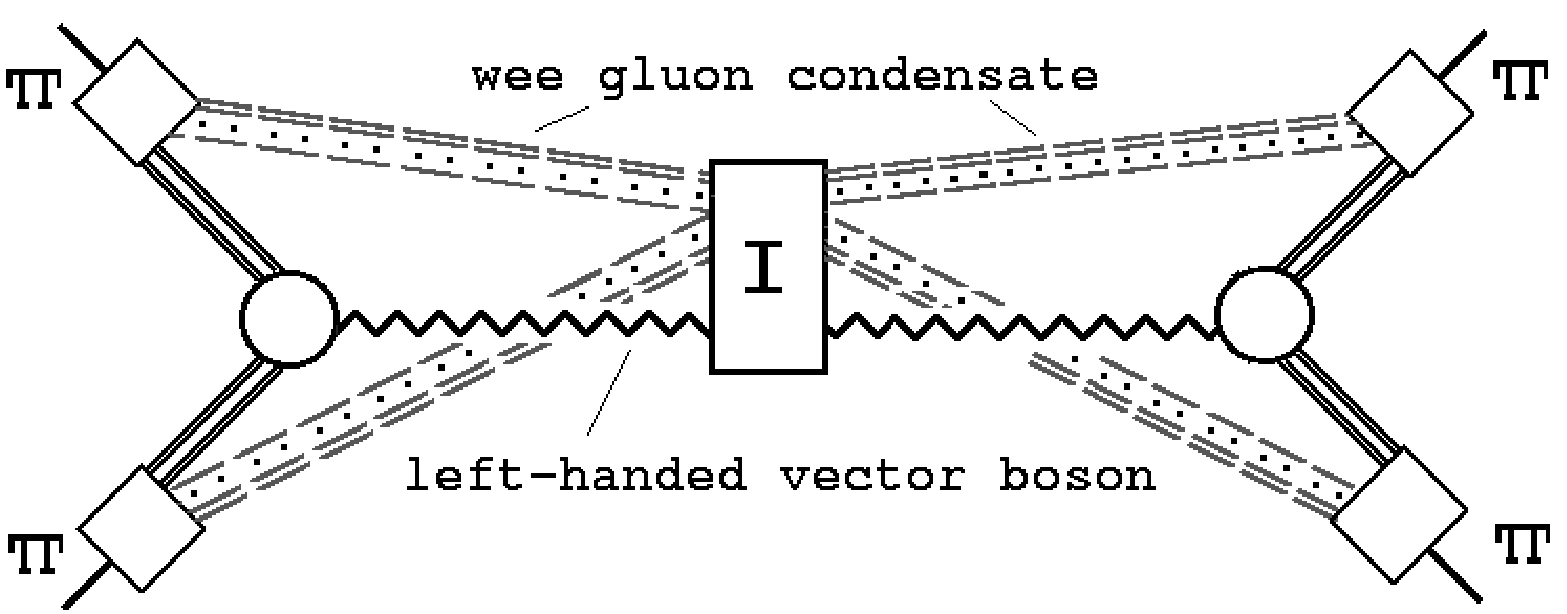}
\hspace{0.1in}
\epsfxsize=2.7in
\epsfbox{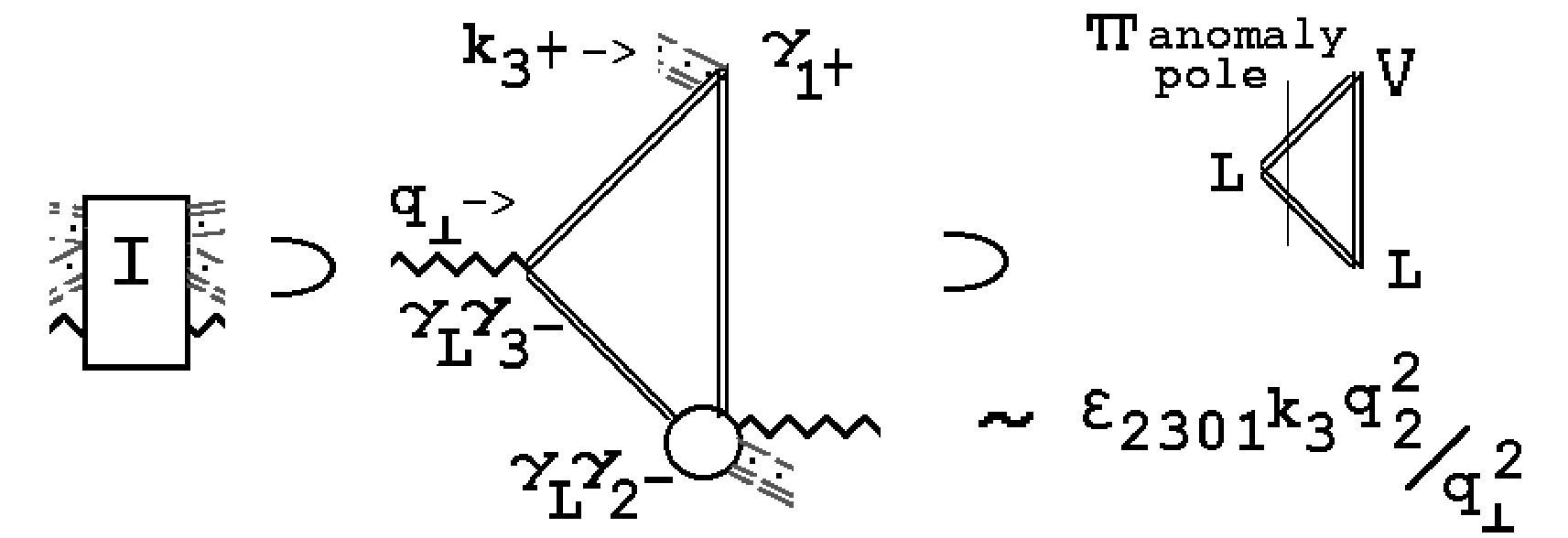}}
\caption{Interactions Producing the Vector Boson Mass.}
\end{figure}
Adding the diagram with $1 \leftrightarrow 2$ 
gives the mass as an integral over wee gluon longitudinal momenta 
($M_W^2 \sim \int k dk $). {\it This mass appears only in the S-Matrix, and occurs only 
for vectors with a left-handed coupling.} 
Both sextet and triplet pions contribute but the 
sextet pions dominate because of larger
color factors. Assuming the Casimir Scaling rule holds
($~ C_6~\alpha_s (F_{\Pi}^2)~\sim ~C_3 ~\alpha_s(F_{\pi}^2)$ with
$C_6/C_3 ~\approx~ 3~$), the sextet chiral scale gives
an electroweak scale of the right magnitude! (The large wee gluon 
coupling to sextet quarks is central in our understanding of high-energy cross-sections.)

\section{Embedding QCD$_S$ in QUD}

If sextet pions are to produce electroweak symmetry breaking, the sextet 
electroweak anomaly must be canceled.
Above the electroweak scale, the $QCD_S$ infra-red fixed point implies that
$\alpha_s$ ${\scriptstyle \leq}$ {\small $1/34$} $ \sim \alpha_{ew}$ 
and so the sextet sector can produce the decrease in $\alpha_s$ 
needed for unification. Supersymmetry is not required !! Looking for a unified theory, 
we found\cite{kw} (a long time ago) a remarkable result. {\it Requiring that
the sextet sector be contained in an asymptotically free, anomaly free, 
theory uniquely selects QUD.}

The SU(3)$\otimes$SU(2)$\otimes$U(1)
decomposition of the individual fermion representations is\cite{kw}
\vspace{0.1in}
\newline {\small
$ ~~~~~$ $5=(1,3,-\frac{1}{3})^{\{3\}}+(2,1,\frac{1}{2})^{\{2\}}, ~~~~~~~~~~~
15=(1,3,1)+(3,2,\frac{1}{6})^{\{1\}}+\{6,1,-\frac{2}{3}\} ~~~~~~~~~~~
$
\newline 
$~~~~$ $ 40=(1,2,-\frac{3}{2})^{\{3\}}+(3,2,\frac{1}{6})^{\{2\}}+
(3^*,1,-\frac{2}{3})+(3^*,3,-\frac{2}{3})+
\{6^*,2,\frac{1}{6}\}~+~(8,1,1)~~~~~~~~~~~
$
\newline
$~~~~$ $45^*=(1,2,-\frac{1}{2})^{\{1\}}+(3^*,1,\frac{1}{3})
+(3^*,3,\frac{1}{3})+(3,1,-\frac{4}{3})+(3,2,\frac{7}{6})^{\{3\}} +
\{6,1,\frac{1}{3}\} +(8,2,-\frac{1}{2}) 
$
\newline $~$}
\newline {\it Very importantly, as we will see, the complete representation is real 
with respect to} SU(3)$\otimes$U(1)$_{em}$. 
There are three ``generations'' of quarks/anti-quarks (labeled ${\scriptstyle
\{1\},\{2\},\{3\}}$)
with quark charges $\frac{2}{3}$ and $-\frac{1}{3}$ and so QUD contains QCD$_S$. 
There are also three ``generations'' of leptons - 
$SU(2)$ doublets that are $SU(3)$ singlets. 
{\it Given that there is no freedom to add more fermions, 
it is very fortunate that we already have the triplet quark and lepton sectors of
the Standard Model, together with dark matter!!} However,  
the $SU(2)\otimes U(1)$ quantum numbers are clearly not quite right. 
Also, at first sight, the color octet quark sector is completely unwanted, as are
the exotic quarks.
 
It would be bizarre indeed if we had arrived at a
unique theory that ``almost'' produces the Standard Model. Originally,
although we saw that QUD has the same asymptotic freedom saturation 
properties as massless QCD$_S$, we did not understand the construction of
high-energy QCD$_S$ well enough to appreciate that high-energy QUD could be 
constructed in a similar manner. Once this is understood, it is only a short step
to the stunning realization that the Standard Model S-Matrix could actually emerge.
 
\section{Construction of QUD High-Energy States and Amplitudes}

We refer to SU(5) gauge boson reggeons as unons. It will be essential that, because
unons are gauge invariant, 
the symmetries we discuss are global and cancelations
can involve unons carrying very different transverse momentum.  
We will identify three fundamental dynamical elements as crucial
in producing the states and S-Matrix of the Standard Model from QUD. 
The first, and most important, is
\medskip
\newline 
$~~$ {\bf [1]} $~$ {\it interactions of left-handed unons 
exponentiate ``anomalous'' divergences.}
\medskip
\newline 
As a result, ``wee unon anomaly divergences''  
and the corresponding dynamical chirality transitions,
can only involve unon combinations within a 
maximal non-abelian vector subgroup. This selects
the strong interaction as involving (a sum over) unon combinations that are 
each singlets under some SU(3) subgroup. To construct high-energy amplitudes explicitly, 
we go to the multi-regge kinematic region, as we did when discussing QCD$_S$.

We start within CSQUD (SU(5) color broken to SU(4)) and expect to find
high-energy QUD as a critical phenomenon. 
Imposing a $k_{\perp}$ cut-off and choosing the vector SU(3)
symmetry as illustrated in Figure 4(a)
we also, initially, break SU(4) to SU(2)$_C$.
The resulting massive unons are an SU(4) singlet vector x,  
two SU(2)$_C$ doublet vectors x', and 
left-handed SU(2)$_C$ singlets x''. 
As illustrated in Figure 4(b), we consider the scattering of x'' unons in the
multi-regge region. (The x'' unons will decouple as SU(5) symmetry 
is restored, leaving only bound-state amplitudes.) 
\begin{figure}[ht]
\centerline{\epsfxsize=1.7in
\epsfbox{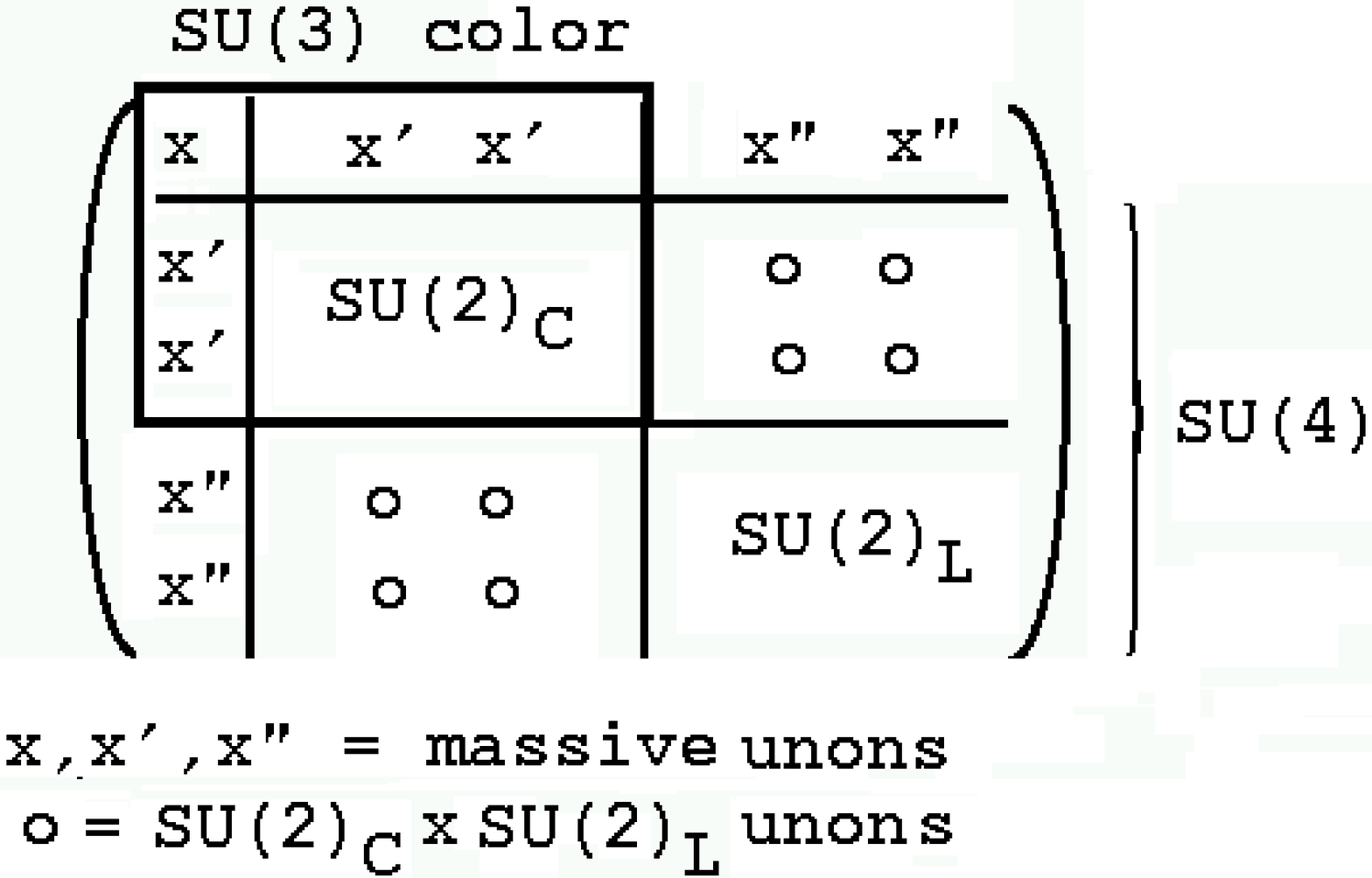}
\hspace{0.3in}
\epsfxsize=1.9in
\epsfbox{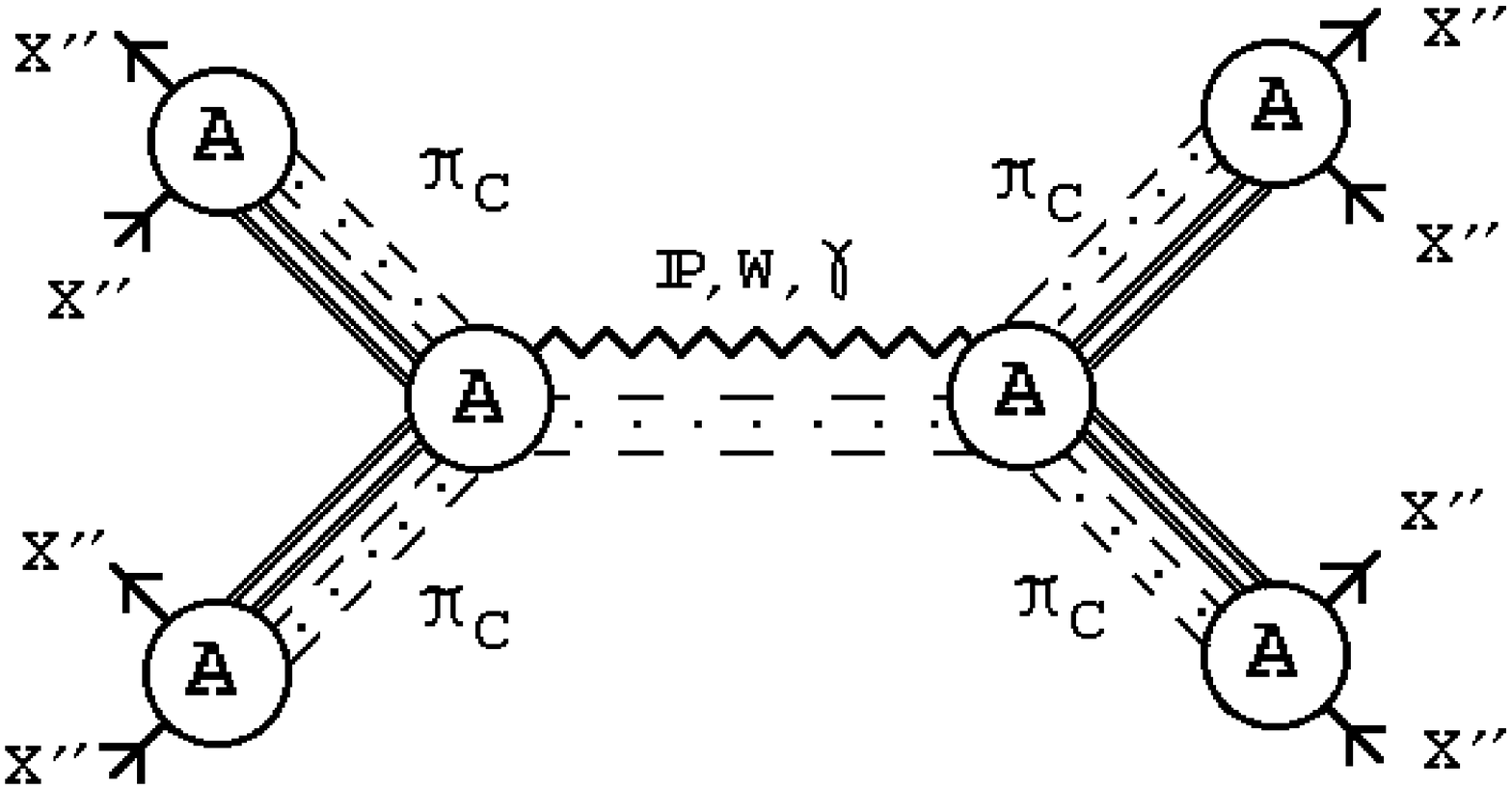}}
\centerline{(a) \hspace{2.1in}(b)}
\caption{(a) Symmetry Breaking (b) Vector Boson Scattering}
\end{figure}

SU(2)$_C$ is a vector symmetry 
and so, as in CSQCD$_S$, there will be a 
``wee gluon condensate'' that produces the physical amplitudes. 
All states are SU(2)$_C$  singlet chiral Goldstone bosons ($\pi_C$'s). 
They are quark, antiquark, and quark/antiquark pairs, 
with one of the pair unphysical (or, equivalently, both physical within the condensate).
Under SU(3) color, the quarks are
{\bf 3's, 6's,} and also {\bf 8's}. The {\bf 8's} appear because 
\medskip
\newline 
$~$ {\bf [2]} {\it octets are real under SU(3),  
but contain complex conjugate doublets under SU(2)$_C$.}
\medskip
\newline 
This is the second fundamental dynamical element. The doublets have the necessary chiral symmetry 
to form Goldstone boson physical states via SU(2)$_C$ divergences.

The x' and SU(2)$_C\otimes$SU(2)$_L$ unons  will be confined via SU(2)$_C$ divergences.
Interactions due to SU(2)$_C$ singlet unons in the wee gluon condensate will be
\medskip{\it \newline $~~~~$ 1. Exchange of a massive x gluon in the condensate 
$\leftrightarrow$ pomeron.
\newline  $~~~~$ 2. Exchange of SU(2)$_L \otimes$ U(1) unons 
in the condensate $\leftrightarrow$ W$^{\pm,0}$,Y.
\newline  $~~~~$ 3. Exchange of a massive x'' unon in the condensate.}
\medskip \newline together with the exchange of any combination.
The `` vacuum properties'' of the SU(2)$_C$ condensate are 
the third fundamental dynamical element. As described in Section 5, 
\medskip
\newline $~~$ {\bf [3]} $~$
{\it wee gluon interactions give left-handed SU(2)$_L \times$ U(1) unons a mass.}
\medskip
\newline 
Only left-handed unons ($W^{\pm}$ and $Z^0$) aquire a mass this way. 
In even signature channels, a similar mixing should
generate masses for all $\pi_C$'s 
that do not couple to the $W^{\pm}$ and $Z^0$. (There are no exact chiral symmetries,
and hence, no massless particles in QUD.)

Because the unons involved are either left-handed or abelian,
restoring SU(2)$_L\otimes$U(1) symmetry gives no new anomaly divergences, but reggeization 
divergences imply that only SU(4) invariant states and interactions survive. 
``Leptons'' are present 
as bound states of elementary leptons and octet pions. For example,
in SU(3)$\otimes$SU(2)$\otimes$U(1) notation,
the electron/neutrino will be {\small
$(1,2,-\frac{1}{2}) \times (8,1,1) \times (8,2,-\frac{1}{2})
\leftrightarrow $} SU(5) singlet {\small 45$^*\times$40$\times$45$^*$}. (The muon 
will contain three elementary leptons, as will the $\tau$.)

Full SU(5) symmetry is achieved by removal of the cut-off, followed by SU(3) restoration.
The pomeron becomes critical and the wee gluon component of the photon and the $W^{\pm}, Z^0$ 
becomes even signature (essentially the zero transverse momentum component of 
the pomeron). The octet pions are no longer Goldstone bosons and so 
they disappear from the low transverse momentum region. 
SU(3) reality also implies they have no anomaly coupling to the pomeron. As a result, 
leptons have no strong interaction and no infra-red SU(3) mass generation.
Because the octet pions contribute only at large transverse 
momentum, the SU(2)$_L\otimes$U(1) symmetry will appear as physical in 
low transverse momentum interactions (with SU(2)$_L \to$  
sextet pion flavor symmetry). Assuming the  
SU(2)$_L\otimes$U(1) anomaly cancellation is maintained by bound-states,
three generations of ``hadrons'' will form from triplet quarks and octet
pions, as low mass bound-state partners of the leptons.
The SU(2)$_L\otimes$U(1) quantum numbers of the octet pions are
$(2,\frac{1}{2}),~ (1,-1)$, and $(3,-1)~$, with unons canceling the 
triplet. Therefore, at low transverse momentum,
the states will have the 
SU(2)$\otimes$U(1) singlet/doublet generation structure of the Standard model. 
{\it Clearly, the octet quarks, which at first sight seem unwanted, 
are fundamental for SU(5) invariance and the generation structure of states.}

We are a long way from establishing much of what we have described and obvious 
questions have been left unanswered (partly because of lack of space).  
There is also much more that we have not discussed at all. Nevertheless, the
possibility that QUD produces the Standard Model S-Matrix seems very real indeed.

\end{document}